\documentclass[10pt,conference]{IEEEtran}
\usepackage{graphicx}
\usepackage[utf8]{inputenc}
\usepackage{mathtools}

\let\subparagraph\relax
\usepackage{subfiles}
\usepackage[justification=centering,font=scriptsize,skip=1pt, belowskip=1pt,aboveskip=1pt]{caption}
\usepackage{color}
\usepackage{titlesec}
\usepackage{amsmath}
\usepackage{pgfplots}
\usepackage{xspace}
\usepackage{authblk}

\let\OLDthebibliography\thebibliography
\renewcommand\thebibliography[1]{
  \OLDthebibliography{#1}
  \setlength{\parskip}{0pt}
  \setlength{\itemsep}{0pt plus 0.0ex}
}
\expandafter\def\expandafter\normalsize\expandafter
{%
    \normalsize
    \setlength\abovedisplayskip{2pt}
    \setlength\belowdisplayskip{0pt}
    \setlength\abovedisplayshortskip{0pt}
    \setlength\belowdisplayshortskip{0pt}
}

\titlespacing\section{0pt}{2pt plus 4pt minus 2pt}{2pt plus 2pt minus 2pt}
\titlespacing\subsection{0pt}{2pt plus 4pt minus 2pt}{2pt plus 2pt minus 2pt}
\titlespacing\subsubsection{0pt}{2pt plus 4pt minus 2pt}{2pt plus 2pt minus 2pt}

\usepackage{listings} 
\usepackage{xcolor}

\definecolor{codegreen}{rgb}{0,0.6,0}
\definecolor{codegray}{rgb}{0.5,0.5,0.5}
\definecolor{codepurple}{rgb}{0.58,0,0.82}
\definecolor{backcolour}{rgb}{0.95,0.95,0.92}
\lstdefinestyle{mystyle}{
    backgroundcolor=\color{backcolour},   
    commentstyle=\color{codegreen},
    keywordstyle=\color{magenta},
    numberstyle=\tiny\color{codegray},
    stringstyle=\color{codepurple},
    basicstyle=\ttfamily\footnotesize,
    breakatwhitespace=false,         
    breaklines=true,                 
    captionpos=b,                    
    keepspaces=true,                 
    showspaces=false,                
    showstringspaces=false,
    showtabs=false,                  
    tabsize=2
}

\newcommand{\sksim}{\textit{SkipSim}\xspace}

\begin{document}

\title{\sksim: Scalable Skip Graph Simulator}
\author{
Yahya Hassanzadeh-Nazarabadi\IEEEauthorrefmark{1}\IEEEauthorrefmark{2}, Ali Utkan Şahin\IEEEauthorrefmark{2},
Öznur Özkasap\IEEEauthorrefmark{2}, 
and Alptekin Küpçü \IEEEauthorrefmark{2}
\\DapperLabs, Vancouver, Canada\IEEEauthorrefmark{1}
\\Department of Computer Engineering, Koç University, İstanbul, Turkey\IEEEauthorrefmark{2}\\
{\{yhassanzadeh13, asahin17, oozkasap, akupcu\}}@ku.edu.tr}

\maketitle
\begin{abstract}
\sksim is an offline Skip Graph simulator that enables Skip Graph-based algorithms including blockchains and P2P cloud storage to be simulated, while preserving their scalability and decentralized nature. To the best of our knowledge, it is the first Skip Graph simulator that provides several features for experimentation on Skip Graph-based overlay networks. In this demo paper, we present \sksim features, its architecture, as well as a sample blockchain demo scenario. 

\end{abstract}

\begin{IEEEkeywords}
Skip Graph, Simulation, Cloud Storage, Blockchain.
\end{IEEEkeywords}

\section{Introduction}
As a Distributed Hash Table (DHT), Skip Graphs \cite{aspnes2007skip} support an extensive domain of applications as part of the P2P cloud storage  \cite{hassanzadeh2016laras, hassanzadeh2019decentralized} and blockchain \cite{hassanzadeh2019lightchain} systems.  
For the design, implementation, and evaluation of the distributed Skip Graph-based protocols, we developed \sksim \cite{skipsim}, which is the first Skip Graph simulator consisting of the following list of \sksim features. 

\textbf{Decentralized:} \sksim preserves the decentralized characteristics of the Skip Graphs by providing and maintaining a local view of the system for each individual node as well as a message-passing communication channel among the nodes.

\textbf{Offline:} In contrast to the online testbeds (e.g., PlanetLab \cite{chun2003planetlab}) that leave the nodes' life cycle management and maintenance on their users, \textit{SkipSim} spawns and manages large scale of nodes in an offline manner and on the same machine. \textit{SkipSim} minimizes the user interaction solely to the instantiation of the simulations and collecting the simulation results, while it handles all the life cycles of the nodes and simulations.

\textbf{Scalable:} On a $2.3$ GHz $8$-Core Intel Core $i9$ with $8$ GB of memory, a fully functional P2P cloud storage system simulation scenario of SkipSim scales up the nodes to around $4000$. Considering the blockchain simulation module, however, \sksim is able to scale up its simulation size to around $10K$ by lightening some unnecessary modules.    

\textbf{Heterogeneity Support:} Peers in the P2P systems are heterogeneous with respect to their resources, e.g., storage capacity and bandwidth. \sksim offers embedded resource distributions of the nodes that are extracted from the traces of real P2P systems \cite{pouwelse2004measurement}. 

\textbf{Churn Support:} Churn is one of the fundamental traits of P2P systems that is defined as the dynamic arrivals and departures of nodes to and from the system, respectively. \sksim comes with several embedded churn models that are extracted from the churn traces of the real P2P systems \cite{stutzbach2006understanding}. 

\textbf{Multi-topology:} To thwart the simulation noises that ground in the biased distribution of the placement of nodes in the underlying network, \sksim spans the simulation configuration over multiple randomized topologies. By the randomized topologies, we mean that the nodes' resources as well as their placement in the underlying network are generated based on specific embedded randomized distributions. Hence, \sksim simulates the same configuration over several topologies and aggregates and outputs the results.

\textbf{Stateful:} In contrast to the similar simulators (e.g., VIBES \cite{stoykov2017vibes}), \sksim captures and records the state snapshot of the systems during the simulation. The snapshots are recorded and collected on a per topology basis. Each snapshot corresponds to a single simulation step, and contains \textit{executable} logs like the churn of the system as well as the query interactions among the nodes. This enables the simulation over each topology to be replayed over a different set of configurations. In other words, \sksim enables the user to replay the same simulation with different configuration parameters, and hence to perceive the effect of each parameter in isolation. 

\textbf{Libraries:} In addition to the DHT-based blockchain \cite{hassanzadeh2019lightchain}, \sksim provides several implementations of P2P protocols in the contexts of low-latency overlays \cite{hassanzadeh2018decentralized, hassanzadeh2015locality}, high availability overlays \cite{hassanzadeh2019interlaced}, replications \cite{hassanzadeh2019decentralized, hassanzadeh2016awake}, and aggregations \cite{hassanzadeh2017elats}.

\begin{figure}
\centering
\includegraphics[width=\linewidth]{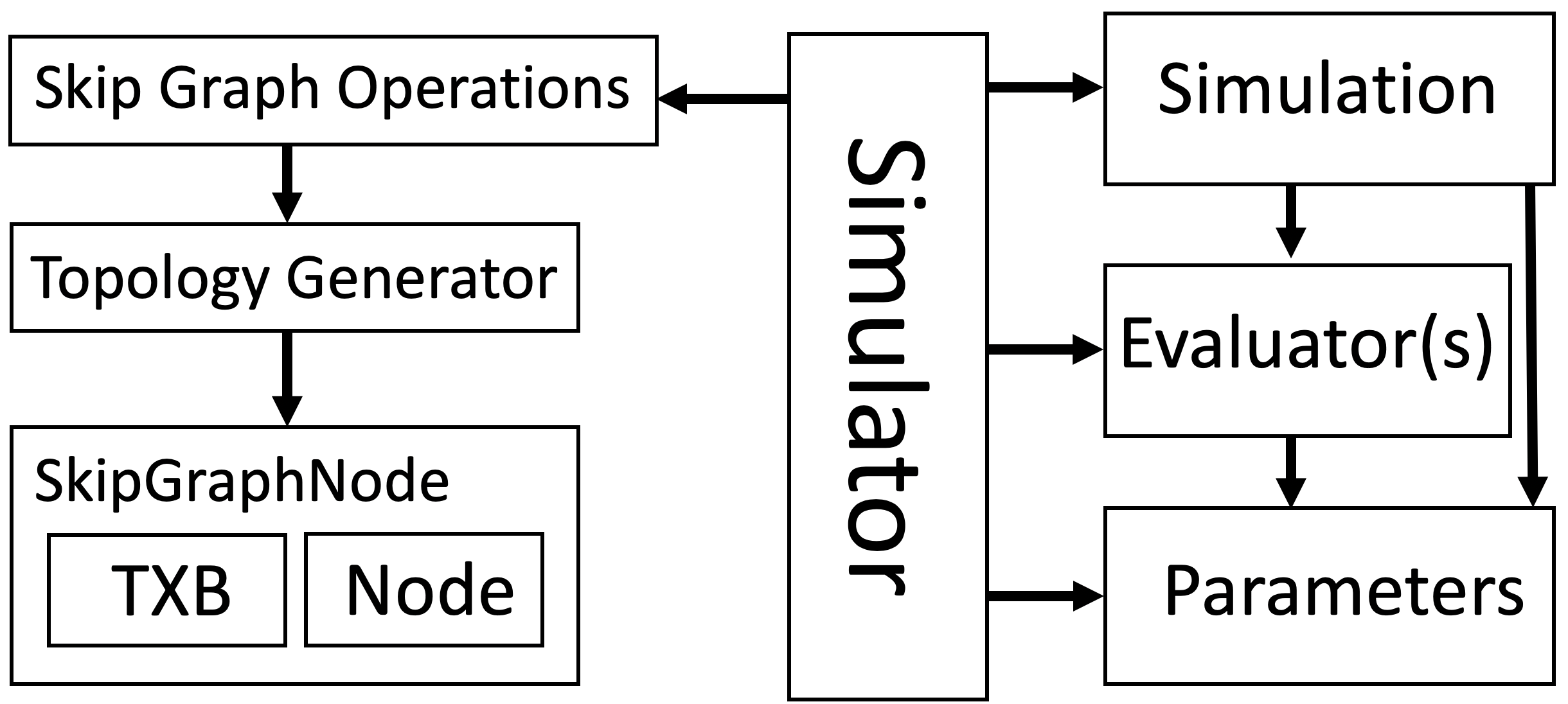}
\caption{An overview on main classes of \textit{SkipSim}}
\label{fig:skipsim}
\end{figure}

\section{Software Architecture}
\textbf{Nodes:} Figure \ref{fig:skipsim} illustrates the architectural components of the \sksim and their interactions. Each rectangle in this figure indicates a \sksim component that is implemented as a Java class. The interactions between classes are represented by arrows: the class on the tail of an arrow has a field of the type of the class on the head of the arrow. The inheritances among the classes are represented by enclosing the children classes into their parents. The \texttt{SkipGraphNode} is the fundamental model class of the \sksim that represents an individual Skip Graph node in an abstract form. Each Skip Graph node is represented by an object of the \texttt{SkipGraphNode}, which contains the basic characteristics of a node such as its identifier, routing (i.e., lookup) table, as well as a unique index that is used to access the node within the simulator. \sksim supports two types of Skip Graph nodes; physical peers and data objects. A physical peer corresponds to a process that joins the Skip Graph overlay for example as a P2P cloud storage \cite{hassanzadeh2016laras} or blockchain node \cite{hassanzadeh2019lightchain}. Each physical peer in \sksim is represented by an object of the \texttt{Node} class, which inherits from \texttt{SkipGraphNode}. A data object can be a blockchain's block, a transaction, or a cloud storage object. Each data object in \sksim is represented by an object of the type \texttt{TXB}, which inherits from \texttt{SkipGraphNode}. 

\textbf{Topology:} In \sksim, a topology determines the connection links between each node and a subset of other nodes in the system known as the node's neighbors. The \texttt{TopologyGenerator} class creates, maintains, and restores the topology objects as well as the churn of the nodes associated with each topology object.

\textbf{Skip Graph Overlay:} The \texttt{SkipGraphOperations} represents the Skip Graph's operations on the nodes of a topology, e.g., distributed put and get queries \cite{aspnes2007skip}. More specifically, this class acts as a communication medium among the nodes by receiving a query from the node that initiates it, routing the query among the other nodes of Skip Graph, and returning the results to the initiator.  

\textbf{Simulation and Evaluation:} The main class of \sksim is the \texttt{Simulator} class, which is responsible for presenting a user interface, loading the simulation parameters, and creating and executing a \texttt{Simulation} object. To be simulated, each algorithm of interest is implemented by extending a \texttt{Simulation} object. Examples of such algorithms currently supported by \sksim are blockchain \cite{hassanzadeh2019lightchain}, replication \cite{hassanzadeh2016laras}, identifier assignment \cite{hassanzadeh2015locality}, fault tolerance \cite{hassanzadeh2019interlaced}, and aggregation \cite{hassanzadeh2017elats}. Each \texttt{Simulation} object is also in interaction with one or more \texttt{Evaluator} objects, which measure the performance of the simulation based on some metrics of interest. Examples of such evaluation metrics are the success probability of adversarial nodes on forking the blockchain \cite{hassanzadeh2019lightchain}, data objects availability \cite{hassanzadeh2016awake}, query latency \cite{hassanzadeh2016laras}, and query success ratio under churn \cite{hassanzadeh2019interlaced}.
\begin{lstlisting}[language=Java, caption= A sample blockchain simulation schema in \sksim, label=simschema]
public class Blockchain extends Parameters
{
    public Blockchain()
    {
        SimulationType = BLOCKCHAIN;
        Blockchain.Protocol = LIGHTCHAIN;
        Topologies = 100;
        SystemCapacity = 1024;
        LifeTime = 168;
        TXB_RATE = 1;
        ChurnModel = FAST_DEBIAN;
        ChurnType = ADVERSARIAL;
        Malicious = 0.16 
        LOG = true;
    }
}
\end{lstlisting}

\section{Sample Demo Scenario}
Listing \ref{simschema} represents a sample blockchain demo configuration in \sksim. A simulation configuration is done by extending the \texttt{Parameters} class of \sksim. During this demo, \sksim spawns $100$ topologies, each with $1024$ nodes, where each node runs an instance of the LightChain blockchain protocol \cite{hassanzadeh2019lightchain}. For each topology, the simulation has a \texttt{Lifetime} of $168$ timeslots (i.e., steps). The nodes in each topology follow a \texttt{FAST$\_$DEBIAN} \cite{hassanzadeh2019decentralized} churn model to imitate online and offline behavior. Each online node creates a single transaction (i.e., \texttt{TXB$\_$RATE $=1$}) at each timeslot. A fraction of $0.16$ of the nodes in the system in this demo are \texttt{Malicious}, i.e., they do not follow the protocol under simulation, i.e., LightChain. The \texttt{ChurnType} parameter is set to \texttt{ADVERSARIAL}, in which peers abruptly leave the network without informing their neighbors. The alternative for the \texttt{ChurnType} is \texttt{COOPERATIVE} in which the peers inform their neighbors on their departure from the system, and hence maintain the overlay connectivity. Also, the current version of \sksim has three churn models that are based on traces of real systems, i.e., \texttt{FAST$\_$DEBIAN}, \texttt{SLOW$\_$DEBIAN}, and \texttt{FLATOUT} \cite{hassanzadeh2019decentralized}.  

After a simulation is configured, the simulation can be started by compiling the source files together. Upon execution \sksim checks its database against the existence of a configuration with the same number of nodes, topologies, churn model and lifetime. In case the same configuration exists, \sksim replays the same snapshot logs on the protocol of interest. Otherwise, it generates and stores a snapshot log for each time slot of the system and plays that log on the protocol of interest. The examples of results extracted from similar demo configurations to Listing \ref{simschema} is available in \cite{hassanzadeh2019lightchain}.

\bibliographystyle{IEEEtran}
\bibliography{references}
\end{document}